\documentclass[12pt]{article}%
\usepackage{amsmath}
\usepackage{amsfonts}
\usepackage{amssymb}
\usepackage{graphicx}%
\providecommand{\U}[1]{\protect\rule{.1in}{.1in}}

\newtheorem{remark}{Remark}

\begin{document}

\title{Energy exchange in Weyl geometry}
\author{John Miritzis}
\date{\today}
\maketitle

\begin{abstract}
We study homogeneous and isotropic cosmologies in a Weyl spacetime. We show
that the field equations can be reduced to the Einstein equations with a
two-fluid source and analyze the qualitative, asymptotic behavior of the
models. Assuming an interaction of the two fluids we impose conditions so that
the solutions of the corresponding dynamical system remain in the physically
acceptable phase space. We show that in Weyl integrable spacetime, the
corresponding scalar field acts as a phantom field and therefore, it may give
rise to a late accelerated expansion of the Universe.

\end{abstract}

\section{Introduction}

The usual approaches for an explanation of the late-time acceleration of the
universe are characterized by a departure from conventional cosmology. The
proposed models, either assume the existence of dark energy \cite{sast,pera},
or require a modification of general relativity at cosmological distance
scales \cite{cdtt,chib}, (cf. \cite{cst,sofa,fets,bcno} for comprehensive
reviews and references). Less explored is the idea that the geometry of
spacetime is not the so far assumed Lorentz geometry (see for example
\cite{cct}). Due to its simplicity Weyl geometry is considered as the most
natural candidate for extending the Lorentzian structure.

We recall that a Weyl space is a manifold endowed with a metric $\mathbf{g}$
and a linear symmetric connection $\mathbf{\nabla}$ which are interrelated
via
\begin{equation}
\nabla_{\mu}g_{\alpha\beta}=-Q_{\mu}g_{\alpha\beta},\label{weyl}%
\end{equation}
where the 1-form $Q_{\mu}$ is customarily called Weyl covariant vector field
(see the Appendix in \cite{miri1} for a detailed exposition of the techniques
involved in Weyl geometry). We denote by $D$ the Levi-Civita connection of the
metric $g_{\alpha\beta}.$

A consistent way to incorporate an arbitrary connection into the dynamics of a
gravity theory is the so-called constrained variational principle \cite{cmq}.
Applying this method in the context of Weyl geometry to the Lagrangian $L=R,$
one obtains the field equations%
\[
G_{\left(  \mu\nu\right)  }=-\nabla_{\left(  \mu\right.  }Q_{\left.
\nu\right)  }+Q_{\mu}Q_{\nu}+g_{\mu\nu}\left(  \nabla^{\alpha}Q_{\alpha
}-Q^{\alpha}Q_{\alpha}\right)  =:M_{\mu\nu},
\]
where $G_{\left(  \mu\nu\right)  }$ is the symmetric part of the Einstein
tensor (see equations (30) and (31) in \cite{cmq}). If we express the tensors
$G_{\left(  \mu\nu\right)  }$ and $M_{\mu\nu}$ in terms of the quantities
formed with the Levi-Civita connection $D$, the field equations become
\begin{equation}
\overset{\circ}{G}_{\mu\nu}=\frac{3}{2}\left(  Q_{\mu}Q_{\nu}-\frac{1}{2}%
Q^{2}g_{\mu\nu}\right)  .\label{field1}%
\end{equation}
In the case of integrable Weyl geometry, i.e., when $Q_{\mu}=\partial_{\mu
}\phi,$ the source term is that of a massless scalar field. Taking the
divergence of (\ref{field1}) and using the Bianchi identities we conclude
that
\[
D^{\mu}Q_{\mu}=0.
\]

In this paper we study Friedmann-Robertson-Walker (FRW) cosmologies in a Weyl
framework. In Section 2 we explore the field equations derived from the
Lagrangian $L=R+L_{m},$ where the matter Lagrangian, $L_{m},$ is chosen so
that ordinary matter is described by a perfect fluid. It is shown that the
presence of the Weyl vector field can be interpreted as a fluid and we analyze
the asymptotic behavior of the models. Assuming an energy exchange between the
two fluids we extend previous work \cite{miri1}. In Section 3 we consider a
modification of the Einstein-Hilbert Lagrangian, cf. (\ref{action}), which may
provide a mechanism of accelerating expansion.

\section{Interacting fluids}

In the following we assume an initially expanding FRW universe with expansion
scale factor $a\left(  t\right)  $ and Hubble function $H=\dot{a}/a$. We adopt
the metric and curvature conventions of \cite{wael}. An overdot denotes
differentiation with respect to time $t,$ and units have been chosen so that
$c=1=8\pi G.$ Ordinary matter is described by a perfect fluid with
energy-momentum tensor,
\begin{equation}
T_{\mu\nu}=\left(  \rho_{2}+p_{2}\right)  u_{\mu}u_{\nu}+p_{2}g_{\mu\nu
},\label{pefl}%
\end{equation}
supplemented with an equation of state $p_{2}=\left(  \gamma_{2}-1\right)
\rho_{2}$. Since for spatially homogeneous and isotropic spacetimes there is
no preferred direction, $Q^{\mu}$ must be proportional to the fluid velocity
$u^{\mu}$, i.e.,
\[
Q^{\mu}=:qu^{\mu},\;\;\;Q^{2}=Q_{\mu}Q^{\mu}=-q^{2}.
\]
(In vacuum, we have to make the assumption that $Q^{\mu}$ is hypersurface
orthogonal, i.e. it is proportional to the unit timelike vector field which is
orthogonal to the homogeneous hypersurfaces). Formally the right-hand side of
(\ref{field1}) can be rewritten as
\begin{equation}
\frac{3}{2}\left(  Q_{\mu}Q_{\nu}-\frac{1}{2}Q^{2}g_{\mu\nu}\right)  =\left(
\rho_{1}+p_{1}\right)  u_{\mu}u_{\nu}+p_{1}g_{\mu\nu},\label{field2}%
\end{equation}
with
\begin{equation}
\rho_{1}=p_{1}=\frac{3}{4}q^{2},\label{ropi}%
\end{equation}
i.e., the equation of state of the $q-$fluid corresponds to stiff matter.
Therefore we are dealing with a two-fluid model with total energy density and
pressure given by
\begin{equation}
\rho=\rho_{1}+\rho_{2},\;\;\;p=p_{1}+p_{2},\label{rop2}%
\end{equation}
respectively, where
\begin{equation}
p_{1}=\rho_{1},\;\;p_{2}=\left(  \gamma_{2}-1\right)  \rho_{2},\label{state}%
\end{equation}
i.e., $\gamma_{1}=2$ and $\gamma_{2}<\gamma_{1}.$

The field equations are the Friedmann equation
\begin{equation}
H^{2}+\frac{k}{a^{2}}=\frac{1}{3}\left(  \rho_{1}+\rho_{2}\right)
,\label{frie1}%
\end{equation}
and the Raychaudhuri equation%
\begin{equation}
\dot{H}=-H^{2}-\frac{1}{6}\left[  \left(  3\gamma_{1}-2\right)  \rho
_{1}+\left(  3\gamma_{2}-2\right)  \rho_{2}\right]  .\label{rayc1}%
\end{equation}

The Bianchi identities imply that the total energy-momentum tensor is
conserved, so that an interaction between the two fluids is induced. It is
necessary to make an assumption about the interaction between the two fluids
(cf \cite{cowajm}), otherwise the field equations constitute an
underdetermined system of differential equations. The simplest assumption is
that the energy-momentum of each fluid is separately conserved, so that the
two fluids do not interact and the densities decay independently,%
\begin{equation}
\dot{\rho}_{1}=-3\gamma_{1}H\rho_{1},\ \ \ \ \dot{\rho}_{2}=-3\gamma_{2}%
H\rho_{2}.\label{cons}%
\end{equation}
The dynamical system (\ref{frie1})-(\ref{cons}) was analyzed in \cite{miri1}.
It was found that in expanding models the \textquotedblleft
real\textquotedblright\ fluid always dominates at late times and therefore the
contribution of the Weyl fluid to the total energy-momentum tensor is
important only at early times. The purpose of this section is to weaken the
requirement of separate conservation of the two fluids.

In many cosmological situations the transfer of energy between two fluids is
important, so one may assume that the two fluids exchange energy. The
following simple model was proposed by Barrow and Clifton (see \cite{bacli}
for motivation and further examples),
\begin{equation}
\dot{\rho}_{1}=-3\gamma_{1}H\rho_{1}-\beta H\rho_{1}+\alpha H\rho
_{2},\label{cons1}%
\end{equation}%
\begin{equation}
\dot{\rho}_{2}=-3\gamma_{2}H\rho_{2}+\beta H\rho_{1}-\alpha H\rho
_{2},\label{cons2}%
\end{equation}
where $\alpha$ and $\beta$ are constants so that the total energy is conserved
(see also \cite{coki} for a singularity analysis of the master equation
derived in \cite{bacli}).

\begin{remark}
In the case of separately conserved fluids, equations (\ref{cons}) imply that
the sets $\rho_{1}=0$ and $\rho_{2}=0,$ are invariant sets for the dynamical
system and by standard arguments, if $\rho_{i}>0,\ i=1,2,$ for some initial
time $t_{0}$, then $\rho_{i}(t)>0$ throughout the solution. This fact can be
made more transparent by the following argument. Assuming that $\gamma
_{1}>\gamma_{2},$ we define the transition variable $\chi\in\lbrack-1,1]$
\begin{equation}
\chi=\frac{\rho_{2}-\rho_{1}}{\rho_{2}+\rho_{1}},\label{tran}%
\end{equation}
which describes which fluid is dominant dynamically \cite{cowajm}. Applying
the conservation equation to $\rho_{1}$ and $\rho_{2}$, one obtains the
evolution equation of the variable $\chi$,%
\begin{equation}
\dot{\chi}=\frac{3}{2}H\left(  \gamma_{1}-\gamma_{2}\right)  \left(
1-\chi^{2}\right)  ,\label{chi}%
\end{equation}
which implies that the sets $\chi=\pm1$ are invariant under the flow of the
dynamical system. Furthermore, the transition variable $\chi$ is bounded, that
is, if initially $\chi\in\lbrack-1,1],$ it remains in that interval for all
$t$. However, the choice (\ref{cons1}) and (\ref{cons2}), has the peculiarity
that the sets $\rho_{1}=0$ and $\rho_{2}=0,$ are no longer invariant sets for
the dynamical system and therefore the sign of the functions $\rho_{1}$ and
$\rho_{2}$ is not conserved. This is also reflected to the fact that the
transition variable $\chi$ no longer satisfies (\ref{chi}) and eventually
escapes outside the interval $[-1,1]$, thus exhibiting unphysical behavior.
\end{remark}

In order to circumvent these difficulties, one has to impose further
conditions on $\alpha$ and $\beta$. It turns out that the assumption
$\beta=-\alpha$ in (\ref{cons1}) and (\ref{cons2}) is a sufficient condition
for the boundness of the function $\chi$. With this assumption we adopt the
Coley and Wainwright formalism \cite{cowajm}, for a general model with two
fluids. The state of the system consists of the couple $\left(  \chi
,\Omega\right)  ,$ where $\Omega=\Omega_{1}+\Omega_{2}$ is the total density
parameter, $\Omega=\rho/3H^{2}.$ In order to allow for closed models in our
analysis, we define the compactified density parameter\emph{\ }$\omega,$ (see
\cite{wainjm})
\begin{equation}
\Omega=\frac{1}{\tan^{2}\omega},\label{compadjm}%
\end{equation}
or
\[
\omega=\arctan\left(  \frac{\sqrt{3}H}{\sqrt{\rho}}\right)
,\,\;\;\;\text{with \ \ }-\pi/2\leq\omega\leq\pi/2.
\]
We see that $\omega$ is bounded at the instant of maximum expansion ($H=0$)
and also as $\rho\rightarrow0,$ in ever-expanding models. Finally, defining a
new time variable $\tau$ by
\[
\frac{d\tau}{dt}=\frac{3\left(  \gamma_{1}-\gamma_{2}\right)  }{2}\sqrt
{\frac{\rho}{3}}\frac{1}{\cos\omega},
\]
one obtains the following dynamical system
\begin{align}
\frac{d\omega}{d\tau}  &  =-\frac{1}{2}\left(  b-\chi\right)  \cos2\omega
\cos\omega\nonumber\\
\frac{d\chi}{d\tau}  &  =(\chi_{\alpha}^{2}-\chi^{2})\sin\omega
,\label{2fluidjm}%
\end{align}
where the constant $b$ is
\[
b=\frac{3\left(  \gamma_{1}+\gamma_{2}\right)  -4}{3\left(  \gamma_{1}%
-\gamma_{2}\right)  }>-1.
\]
In our case, we always have $\gamma_{1}=2.$ The parameter $\chi_{\alpha}$
which determines the range of $\chi$ is given by%
\[
\chi_{\alpha}=\sqrt{1-\frac{4\alpha}{3\left(  \gamma_{1}-\gamma_{2}\right)  }%
},\ \ \ \ \ \ \ 0\leq\alpha\leq\frac{3\left(  \gamma_{1}-\gamma_{2}\right)
}{4}.
\]

The phase space of the two-dimensional system (\ref{2fluidjm}) is the closed
rectangle
\[
D=\left[  -\pi/2,\pi/2\right]  \times\left[  -\chi_{\alpha},\chi_{\alpha
}\right]
\]
in the $\omega-\chi$ plane (see Figure 1). Since $0\leq\chi_{\alpha}\leq1,$
the rectangle $D$ is shrinked compared to the phase space in \cite{miri1} and
\cite{wainjm}.

\begin{figure}[h]
\begin{center}
\includegraphics{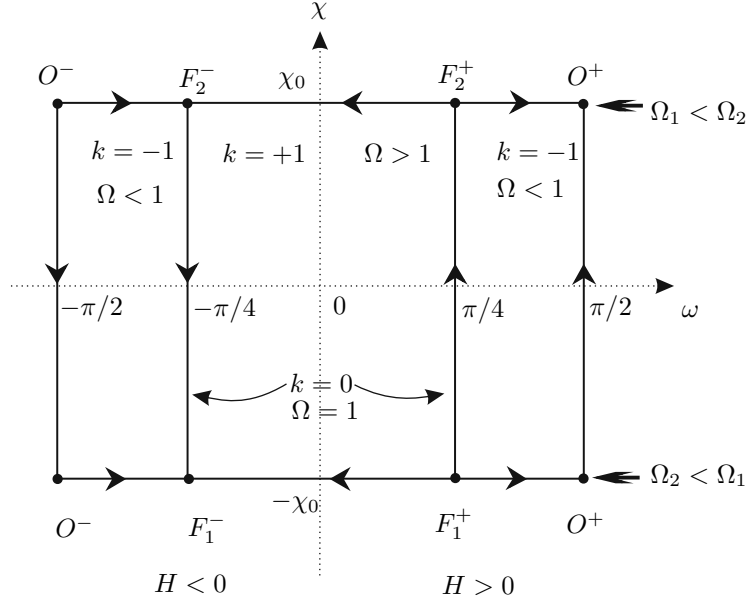}
\end{center}
\par
\caption{The invariant sets and equilibrium points of (\ref{2fluidjm})}%
\label{fig1}%
\end{figure}

The invariant sets of the system are denoted in the following table.%

\[%
\begin{array}
[c]{lll}%
\omega=-\pi/2 & \text{contracting empty models} & \Omega=0,H<0\\
\chi=-\chi_{\alpha} & \text{scaling solution} & \frac{\Omega_{2}}{\Omega_{1}%
}=\frac{1-\chi_{\alpha}}{1+\chi_{\alpha}}\\
\chi=+\chi_{\alpha} & \text{scaling solution} & \frac{\Omega_{1}}{\Omega_{2}%
}=\frac{1-\chi_{\alpha}}{1+\chi_{\alpha}}\\
\omega=\pi/4 & \text{expanding flat models} & \Omega=1,H>0\\
\omega=-\pi/4 & \text{contracting flat models} & \Omega=1,H<0\\
\omega=\pi/2 & \text{expanding empty models} & \Omega=0,H>0
\end{array}
\]
It is easy to verify that the equilibrium points lie at the intersection of
these sets and are denoted by $F_{1,2}^{\pm}$ (expanding or contracting flat
model) and $O^{\pm}$ (expanding or contracting open model). The subscripts
indicate which fluid dominates. Linearization around the equilibrium points is
sufficient for the characterization of their stability and the result is the
phase portrait is shown in Figure 2.

\begin{figure}[h]
\begin{center}
\includegraphics{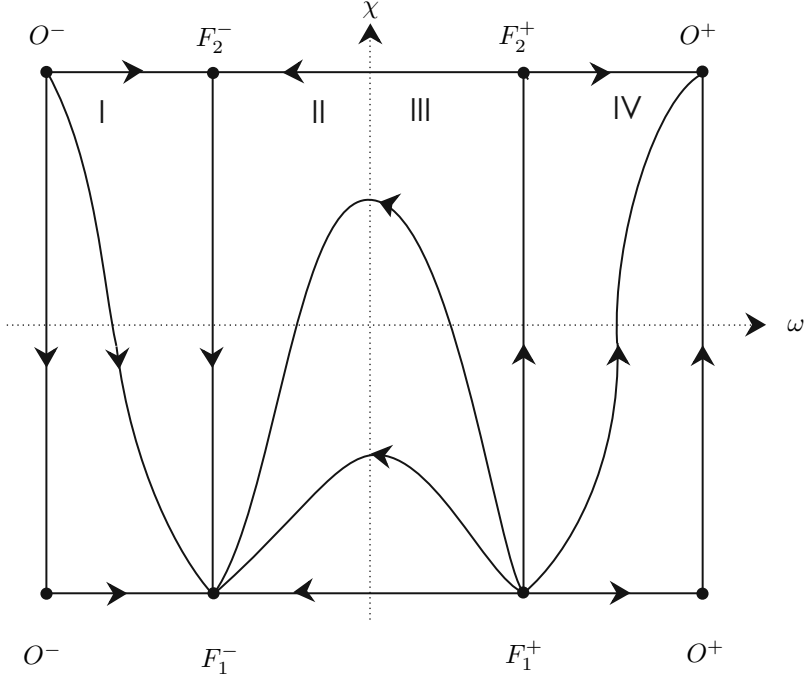}
\end{center}
\caption{The phase portrait of (\ref{2fluidjm}) with $\gamma_{2}=1.$}%
\label{fig2}%
\end{figure}

Regions III and IV correspond to expanding models. The $F_{1}^{+}$ is a past
attractor of all models with $\Omega>0$, i.e., the evolution near the big bang
is approximated by the flat FRW model where the Weyl fluid dominates. Open
models expand indefinitely and approach at late time a \textquotedblleft
scaling solution\textquotedblright\ where the Weyl fluid keeps a small
fraction of the total energy density. Flat models expand indefinitely and the
evolution is approximated by the flat FRW universe at late time. In both cases
the \textquotedblleft real\textquotedblright\ second fluid dominates at late
times. On the other hand, any initially expanding closed model in region III,
however close to $F_{2}^{+}$, eventually recollapses and the evolution is
approximated by the flat FRW model where the Weyl fluid dominates.

We therefore conclude that the Weyl fluid has significant contribution only
near the cosmological singularities. In expanding models the ``real'' fluid
always dominates at late times and therefore the contribution of the Weyl
fluid to the total energy-momentum tensor is important only at early times.

\section{Phantom from pure geometry}

The field equations (\ref{field1}) constitute the generalization of the
Einstein equations in a Weyl spacetime in the sense that they come from the
Lagrangian $L=R.$ There is however an alternative view, namely that the pair
$\left(  Q,\mathbf{g}\right)  $ which defines the Weyl spacetime also enters
into the gravitational theory and therefore, the field $Q$ must be contained
in the Lagrangian independently from $\mathbf{g.}$ In the case of integrable
Weyl geometry, i.e. when $Q_{\mu}=\partial_{\mu}\phi$ where $\phi$ is a scalar
field, the pair $\left(  \phi,g_{\mu\nu}\right)  $ constitute the set of
fundamental geometrical variables. We stress that the nature of the scalar
field is purely geometric. A simple Lagrangian involving the set $\left(
\phi,g_{\mu\nu}\right)  $ is given by
\begin{equation}
L=R+\xi\nabla^{\mu}Q_{\mu}+L_{m},\label{action}%
\end{equation}
where $\xi$ is a constant and $L_{m}$ corresponds to the Lagrangian yielding
the energy-momentum tensor of a perfect fluid. Motivations for considering
theory (\ref{action}) can be found in \cite{nove,sasa} (see also \cite{kome}
for a multidimensional approach and \cite{oss} for an extension of
(\ref{action}) to include an exponential potential function of $\phi$). By
varying the action corresponding to (\ref{action}) with respect to both
$g_{\mu\nu}$ and $\phi$ one obtains
\begin{equation}
\overset{\circ}{G}_{\mu\nu}=\frac{3-4\xi}{2}\left(  \partial_{\mu}\phi
\partial_{\nu}\phi-\frac{1}{2}\left(  \partial_{\alpha}\phi\partial^{\alpha
}\phi\right)  g_{\mu\nu}\right)  +T_{\mu\nu},\label{nove}%
\end{equation}
and%
\begin{equation}
\overset{\circ}{\square}\phi=\frac{1}{3-4\xi}\rho,\label{emsf}%
\end{equation}
where $\overset{\circ}{\square}$ is the D'Alembertian operator formed with the
Levi-Civita connection $D$. As mentioned above, ordinary matter described by
$T_{\mu\nu}$ is a perfect fluid with energy density $\rho$ and pressure $p$.
Setting
\[
\lambda=\frac{4\xi-3}{2},
\]
we note that for $\lambda<0$ the field equations are formally equivalent to
general relativity with a massless scalar field coupled to a perfect fluid. In
Weyl spacetime the scalar field has a geometric nature and no restriction
exists for the sign of the value of $\lambda$. For $\lambda>0,$ the Weyl field
$\phi$ plays the role of a phantom scalar field and therefore, it may provide
a mechanism of late time acceleration. Further investigation of this issue is
the subject of future research.

\end{document}